\documentclass[osajnl,twocolumn,showpacs,superscriptaddress,10pt]{revtex4-1} 
\usepackage{amsmath,amssymb,graphicx}
\usepackage{color}
\def\Ai{\qopname\relax{no}{Ai}}
\def\acos{\qopname\relax{no}{acos}}

\begin{document}

\title{Intermittent Giant Goos-Hanchen shifts from Airy beams at nonlinear interfaces}

\author{Pedro Chamorro-Posada}
\author{Julio S\'anchez-Curto}
\affiliation{Departamento de Teor\'{\i}a de la Se\~nal y Comunicaciones e Ingenier\'{\i}a Telem\'atica, Universidad de Valladolid, ETSI Telecomunicaci\'on, Paseo Bel\'en 15, 47011 Valladolid, Spain}
\author{Alejandro B. Aceves}
\affiliation{Department of Mathematics, Southern Methodist University, Clements Hall 221 ,  Dallas, TX  75275, USA}
\author{Graham S. McDonald}
\affiliation{Joule Physics Laboratory, School of Computing, Science and Engineering, Materials and Physics Research Centre, University of Salford, Salford M5 4WT, United Kingdom}

\begin{abstract}
We study the giant Goos-Hanchen shift obtained from an Airy beam impinging on a nonlinear interface.  To avoid any angular restriction associated with the paraxial approximation, the analysis is based on the numerical solution of the nonlinear Helmholtz equation.  We report the existence of non-standard intermittent and oscillatory regimes for the nonlinear Goos-Hanchen shifts which can be explained in terms of the competition between the critical coupling to a surface mode of the reflected component of the Airy beam and the soliton emission from the refracted beam component.
\end{abstract}

\maketitle 

Airy beams \cite{berry} have remarkable properties, such as their self-healing capabilities \cite{healing}, that make them very attractive for applications ranging from linear optical communications \cite{gu} to the high-intense nonlinear optical regime \cite{polynkin}. Ideal nondiffracting self-bending optical Airy beams have infinite energy and some sort of truncation is required to obtain solutions that can be used in practice.  Finite energy Airy beams keep the main properties of their ideal counterparts only for a limited propagation distance.  Among the various alternatives that have been put forward as finite energy Airy beams we will use in this work the exponential apodization of the initial beam profile proposed in \cite{siviloglou}.  

The refraction and reflection of Airy beams at linear interfaces were addressed in \cite{chremmos}.  Here, we study the behavior of finite energy Airy beams at a linear-nonlinear interface close to the critical angle for total internal reflection.  In this setup, an enhancement of the Goos-H\"anchen shift (GHS) \cite{goos47} can be obtained due to the coupling to nonlinear surface modes.  This giant or nonlinear GHS was first described in \cite{tomlinson82} for Gaussian beams and later studied analytically, for nonlinear interfaces, using a equivalent-particle description of optical solitons \cite{aceves}.  The use of a Helmholtz theory for optical solitons \cite{chamorro98} and their behavior at interfaces \cite{sanchez07}, with the associated numerical tools \cite{chamorro01}, permitted to address this problem without the angular restrictions imposed by the paraxial approximation and to reveal new related phenomena \cite{sanchez11}.  

The dependence of the critical angle with the nonlinear refractive index results in the existence of a critical intensity for the coupling to the surface mode when  a Gaussian beam impinges on a nonlinear interface at fixed angle of incidence \cite{tomlinson82}.  A similar behavior is also found for the interaction of a soliton with a nonlinear interface \cite{aceves,sanchez11}.  Our input Airy condition has a richer phase and amplitude structure when compared with a Gaussian or a soliton beam.  We report that the coupling of the incident Airy beam to the surface mode can exhibit widely-varying GHS values arising from relatively small changes in the effective incident beam intensity at the interface with the result of oscillatory or even intermittent regimes.  These new effects can display an extreme sensitivity to the transmitted field intensity and could be used for sensing applications.  They can be interpreted as the result of the competition between the critical coupling to the surface mode and the soliton shedding from a refracted component of the beam.  At larger intensities, we find that the simultaneous generation of refracted soliton beams and surface modes can be obtained and that this secondary coupling to a surface wave displays similar features as the first order one.

We consider  a plane boundary  separating two media, $l=1,2$, with refractive indices $n_l^2=n_{0,l}^2+n_{NL,l}^2 $ 
and $n_{NL,l}^2=\gamma_l\left|E\right|^2$.  The nonlinear Helmholtz equation describing the evolution of a TE polarized optical field $E(x,z)=E_0u(x,z)\exp(jkn_0 z) $ can be written, in terms of its complex envelope $u(x,z)$, as \cite{chamorro98,sanchez07} 
\begin{equation}
\kappa \dfrac{\partial ^2 u}{\partial \zeta^2}+j\dfrac{\partial u}{\partial \zeta}+\dfrac{1}{2}\dfrac{\partial ^2 u}{\partial \xi^2}-\dfrac{\Delta_l}{4\kappa}u+\alpha_l|u|^2u=0,\label{eq:nnse}
\end{equation}
where 
\begin{equation}
\Delta_l=1-\left(\dfrac{n_{0l}}{n_0}\right)^2,
\end{equation}
$n_0$ is an arbitrary reference refractive index and
\begin{equation}
4\kappa\alpha_l=\dfrac{\gamma_l E_0^2}{n_o^2}\label{alpha}
\end{equation}
is the ratio between the linear and nonlinear refractive indices \cite{tlm}.  The transverse and longitudinal scalings in $\xi=x/X_0$ and $\zeta=z/Z_0$ are chosen such that $Z_0=kn_0X_0^2$  is the Rayleigh range of a hypothetical input Gaussian amplitude of width $X_0$.  Parameter $\kappa=X_0^2/(2Z_0^2)$ keeps the information about the scalings used and permits to rewrite any particular solution of the normalized equation back in the laboratory coordinates. This nonparaxiality parameter \cite{chamorro98} $\kappa=\left((\lambda_0/n_0)/X_0\right)^2/(8\pi)^2$ will be very small for paraxial propagation situations.  The paraxial approximation amounts to neglecting the first term in \eqref{eq:nnse}, so a Nonlinear Schr\"odinger (NLS) Equation describing the evolution of the optical beam is obtained. Once the nonparaxiality parameter $\kappa$ is fixed, we adjust the nonlinear refractive index mismatch by varying a single parameter $\alpha$.  This could correspond to either a change of the input beam intensity or the nonlinear medium parameters.

Optical solitons are perfectly collimated nonlinear beams and the continuity of their phase across the interface permits to obtain a precise nonlinear Snell's law for solitons of the nonlinear Helmholtz equation \cite{sanchez07} that is a direct analogue of the refraction law for linear plane waves or optical rays.  In fact, the giant GHS exhibited by optical solitons close to the critical angle reveals a behavior close to that predicted by geometrical optics for optical rays \cite{sanchez11}.  The same approach permits to write the corresponding condition for nonlinear plane wave solutions of \eqref{eq:nnse}.  In general, the Snell's law is 
\begin{equation}
n_1\cos\theta_1=n_2\cos\theta_2 \label{snell}
\end{equation}  
where  $\theta_1$ and $\theta_2$ are, respectively, the angles of the incident and refracted wave normals with the interface.  For solitons,  $n_l=n_0\sqrt{1-\Delta_l+2\kappa\alpha_l\eta^2}$ and $\eta$ is the soliton parameter that defines its normalized peak amplitude and inverse transverse width $\sqrt{\alpha_l}\eta$.  For plane waves,  $n_l=n_0\sqrt{1-\Delta_l+4\kappa\alpha_l|u|^2}$, and $|u|$ is the normalized amplitude.

The $l=1$ incidence medium will be linear for all the cases considered in this Letter and we will set its refractive index as the reference value for the whole study.  Therefore, we will use $\Delta_1=0$ and $\alpha_1=0$ in Equation \eqref{eq:nnse}.  The refraction of the incident beam will take place at the planar boundary with a focusing Kerr nonlinear medium $\alpha_2=\alpha>0$, such that the discontinuity in the linear refractive indexes should correspond to (linear) internal refraction $\Delta_2=\Delta>0$.

\begin{figure}[htbp]
\centering
\begin{tabular}{cc}
(a)&(d)\\



\includegraphics[width=.5\columnwidth]{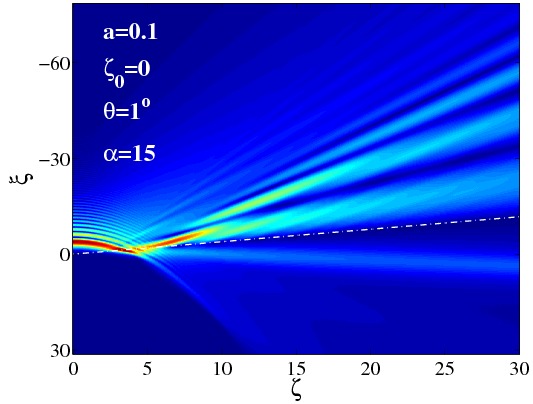}&
\includegraphics[width=.5\columnwidth]{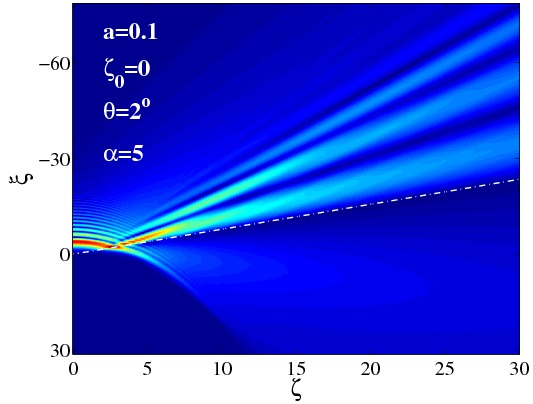}\\

(b)&(e)\\
\includegraphics[width=.5\columnwidth]{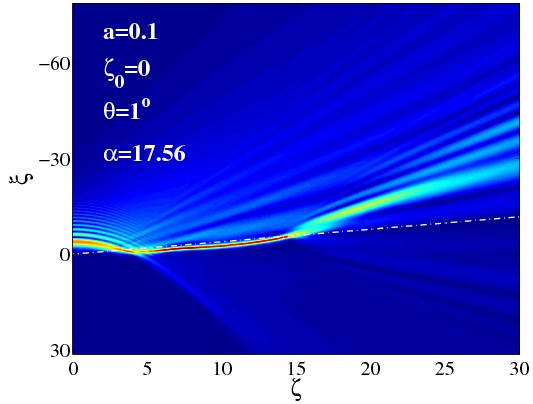}&
\includegraphics[width=.5\columnwidth]{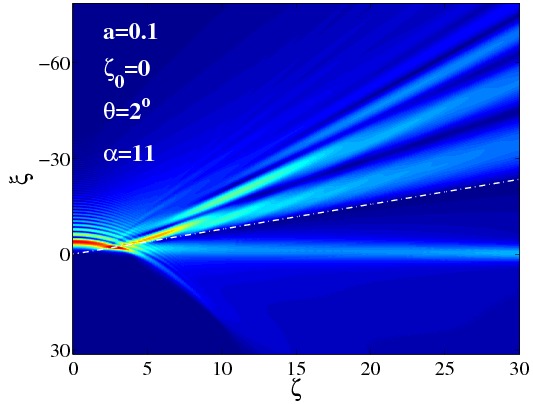}\\

(c)&(f)\\
\includegraphics[width=.5\columnwidth]{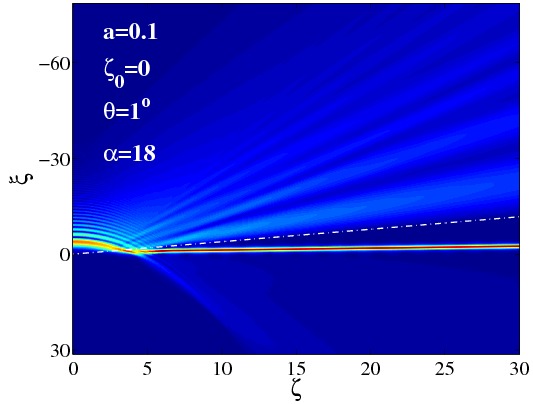}&
\includegraphics[width=.5\columnwidth]{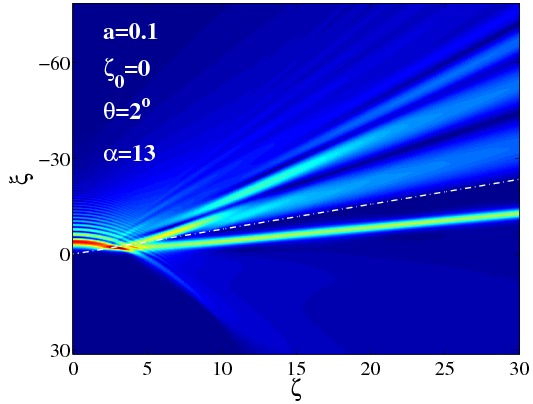}
\end{tabular}
\caption{(a) to (c) Typical sequence in the critical coupling to a surface mode and giant GHS. (d) to (f) Typical soliton shedding sequence as $\alpha$ increases.   In both cases $a=0.1$.}\label{tipicaa01}
\end{figure}

Under these conditions, we can define a critical angle for solitons \cite{sanchez11} $\theta_{c,s}=\acos\sqrt{1-\Delta +2\kappa\alpha \eta^2}$ 
and a corresponding expression for plane waves $\theta_{c,pw}=\acos\sqrt{1-\Delta +4\kappa\alpha |u|^2}$. Both  extreme cases show the same type of dependence of the critical coupling condition on the nonlinear mismatch parameter $\alpha$ that -as we show in this Letter- fails to describe the interaction properties of an input Airy beam with the nonlinear interface.

We will consider an input beam always propagating on-axis, with an initial transverse profile  
\begin{equation}
u(\xi,\zeta=0)=\Ai\left(\xi+\xi_p\right)\exp(a\left(\xi+\xi_p\right))\label{eq:initial}
\end{equation}
that corresponds to the initial condition of a finite energy accelerating Airy beam \cite{siviloglou}. The angular spectrum of this beam has a width of order $1/\sqrt{2a}$ that, as it has been noted in \cite{arxiv}, decreases as its localization at the launch point increases.  

The paraxial description of the initial propagation of the input beam launched along the optical axis in the incidence medium will be adequate provided that $a>>\kappa$ since the maximum transverse wavenumber in the normalized frame \cite{chamorro98} is $1/\sqrt{2\kappa}$. The asymptotic paraxial evolution from the input condition \eqref{eq:initial} admits an approximate Gaussian-like mode that is a very accurate representation of the solution at distances sufficiently far from the launch plane \cite{arxiv} and, therefore, a behavior similar to the corresponding incident Gaussian mode could be expected in this regime.  Therefore, we will focus on interfaces close to the beam launch plane and we will study how the interaction effects change with the distance between the launch point and the interface.

For the analyses, we have used the numerical solution of the full Helmholtz equation \cite{chamorro01}.  This is consistent with the paraxial input conditions, but the validity of the results is also guaranteed for arbitrary angular contents in the reflected and refracted components resulting from the interaction with the interface.  Furthermore, we will use the angular freedom obtained from the Helmholtz framework to rotate the interface an inclination angle $\theta_i$ and keep the input beam fixed in order to change the angle of incidence.  The use of the full Helmholtz framework permits also to unambiguously map an angle $V$ in the normalized frame map to a physical angle through the nonparaxial parameter $\kappa$ as $\theta=\arctan\left(\sqrt{2\kappa} V\right)$ \cite{chamorro98}.

\begin{figure}[htbp]
\centering
\begin{tabular}{cc}
(a)&(b)\\
\includegraphics[width=.5\columnwidth]{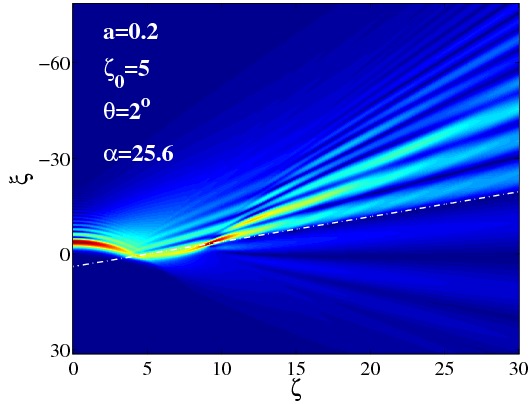}&
\includegraphics[width=.5\columnwidth]{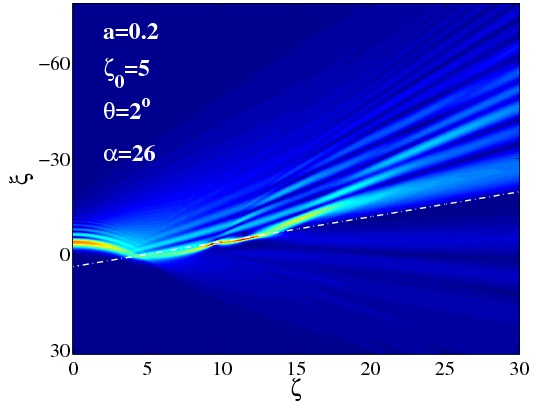}\\
(c)&(d)\\
\includegraphics[width=.5\columnwidth]{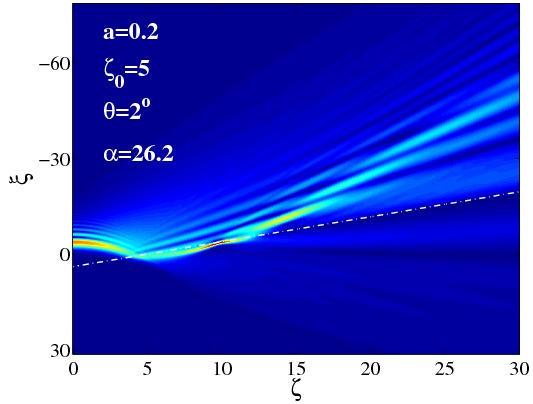}&
\includegraphics[width=.5\columnwidth]{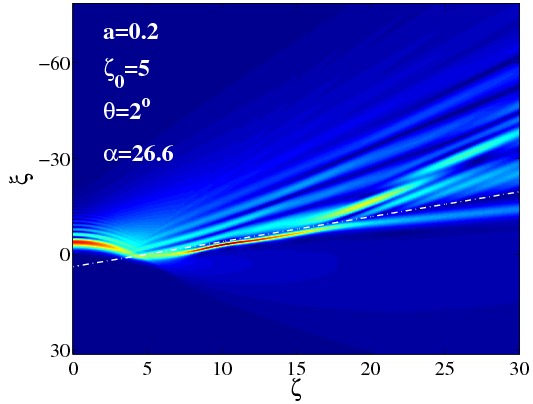}
\end{tabular}
\caption{Oscillations of giant GHS at  $\theta=2^o$, $a=0.2$ and $\zeta_0=5$.}\label{osc}
\end{figure}

\begin{figure}[htbp]
\centering
\begin{tabular}{cc}
\includegraphics[width=.5\columnwidth]{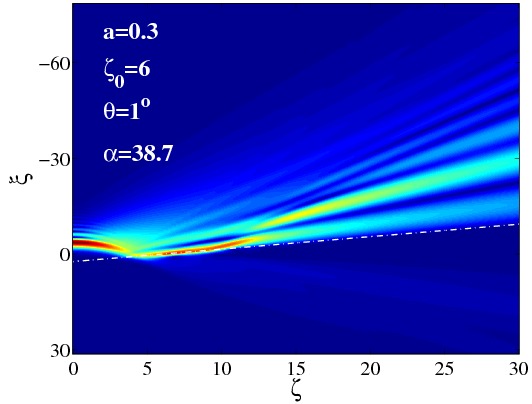}&
\includegraphics[width=.5\columnwidth]{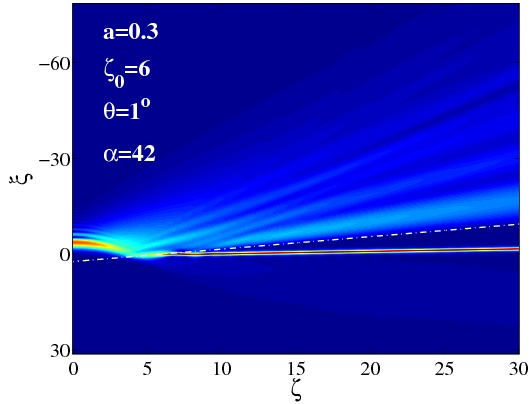}\\
\includegraphics[width=.5\columnwidth]{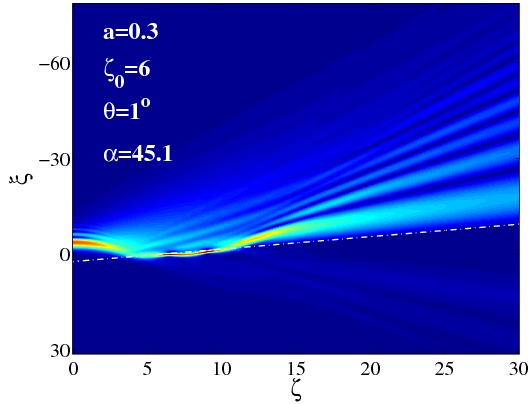}&
\includegraphics[width=.5\columnwidth]{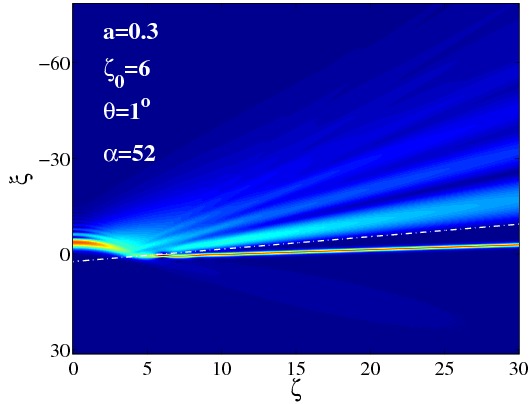}\\
\end{tabular}
\caption{Set-up of an intermittent GHS at $a=0.3$, $\theta=1^o$ $\zeta_0=6$.}\label{int}
\end{figure}

In order to analyze the interaction of the input Airy beam with the interface we have performed extensive numerical studies.   A preliminary survey over a wide range of parameter values permitted to identify the relevant parameter regions that were later studied in a systematic way.  The remainder of this paper highlights unique features that we observed in some regions of parameter space.  The values chosen for parameter $a$ were $0.1$, $0.2$ $0.3$, $0.4$ and $0.6$, ranging over increasing degrees of localization of the input beam.  The input beam was defined with $\xi_p=3$ and the conditions of the incidence on the interface were controlled by pivoting the interface around the point $(\xi_0,\zeta_0)$ with a varying angle $\theta$ (in laboratory coordinates).  Two values were considered for the interface inclination: $\theta=1^o$ and $\theta=2^o$, and $\zeta_0$ took values from $0$ to $12$, always with $\xi_0=0$, and propagation over a total normalized length of $\zeta=30$ was considered in all cases.  For each set of $(a,\zeta_0,\theta)$, $\alpha$ was varied starting from $\alpha=0$; this would correspond, in an associated experimental set-up, to a progressive increase of the input beam power.  We have considered a value of $\Delta=0.01$ and $\kappa=0.001$ in all cases.  

The value of $a$ is central for the behavior of the Airy beam at the interface.  As $a$ decreases the effect of the apodization becomes very small and this affects the possibility of the observation of giant GHS: in our case it is found only when the interface is very close to the launch plane and for the smallest tilt angle, which suggest the existence of a threshold for the observation of GHS.  Also, for $a\ge 0.6$ the oscillations of the Airy beams are largely diminished and the solution is very close to its asymptotic Gaussian beam even at small distances to the launch plane and a behavior close to that of Gaussian input \cite{tomlinson82} is found for most cases.  For intermediate values of $a$ and small $\alpha$, in the linear or quasi-linear regime, the input Airy beam impinging on the interface produces both a refracted and a reflected contribution. 

Two distinct scenarios are described in Figure \ref{tipicaa01}.  Figures (a) to (c) illustrate the evolution from the intial reflection of the main lobe of the input Airy beam as its intensity is varied. The associated GHS grows as $\alpha$ increases towards a critical value, producing giant nonlinear shifts.  When the critical $\alpha$ is exceeded, the soliton propagates fully in the second medium.  The whole picture is very close to that of Gaussian \cite{tomlinson82} or soliton beams \cite{aceves,sanchez11} at nonlinear interfaces.  Nevertheless, the complex structure of Airy beams allows for the existence of a second scenario starting from the reflection of the main lobe of the Airy beam as the intensity increases, as described in Figures (d) to (e).  In this case, the existence of a refracted beam component results in the eventual emission of a soliton in the second medium.    .  

As expected, the soliton shedding tends to dominate at low values of $a$ and the critical GHS at larger values of the beam apodization parameter.  For intermediate values of $a$ the competition of these two effects and the complexity of the structure of the accelerating beam can result in wide variations of GHS with small changes in the input intensity, the existence of oscillations in the observed giant GHS (Fig. \ref{osc}) and intermittency regions (Fig. \ref{int}) for the GHS.  This multiplicity of critical values resembles the coupling to Tamm waves at the interface with a structured material \cite{tamm}.  While there is a boundary separating two media, each is homogeneous; it is then that the unique oscillatory profile of the Airy beam combined with the nonlinear property of the dielectric to the right of the interface induces both the excitement of a surface wave and the Tamm-like effect.

\begin{figure}[htbp]
\centering
\begin{tabular}{cc}
(a)&(b)\\
\includegraphics[width=.5\columnwidth]{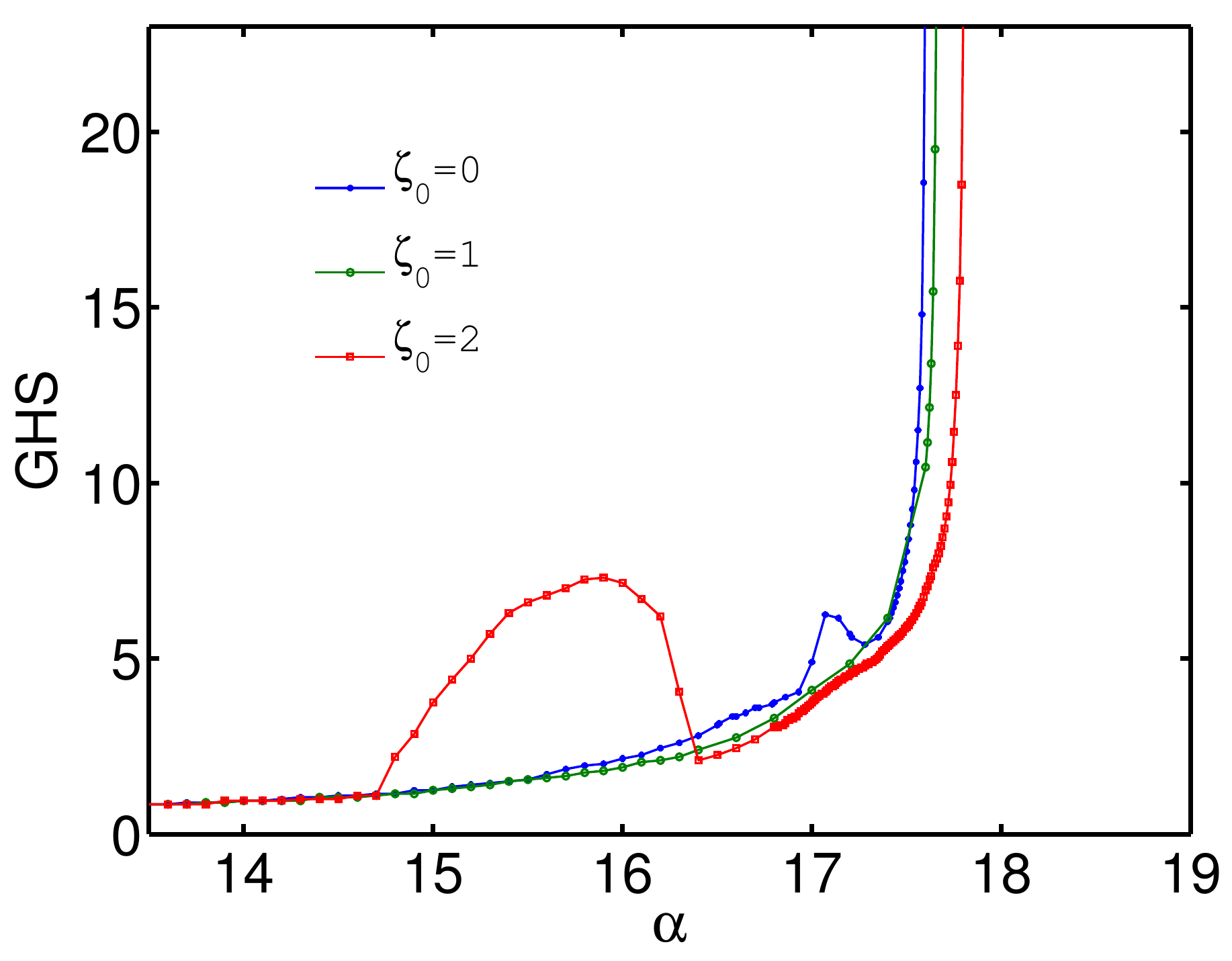}&
\includegraphics[width=.5\columnwidth]{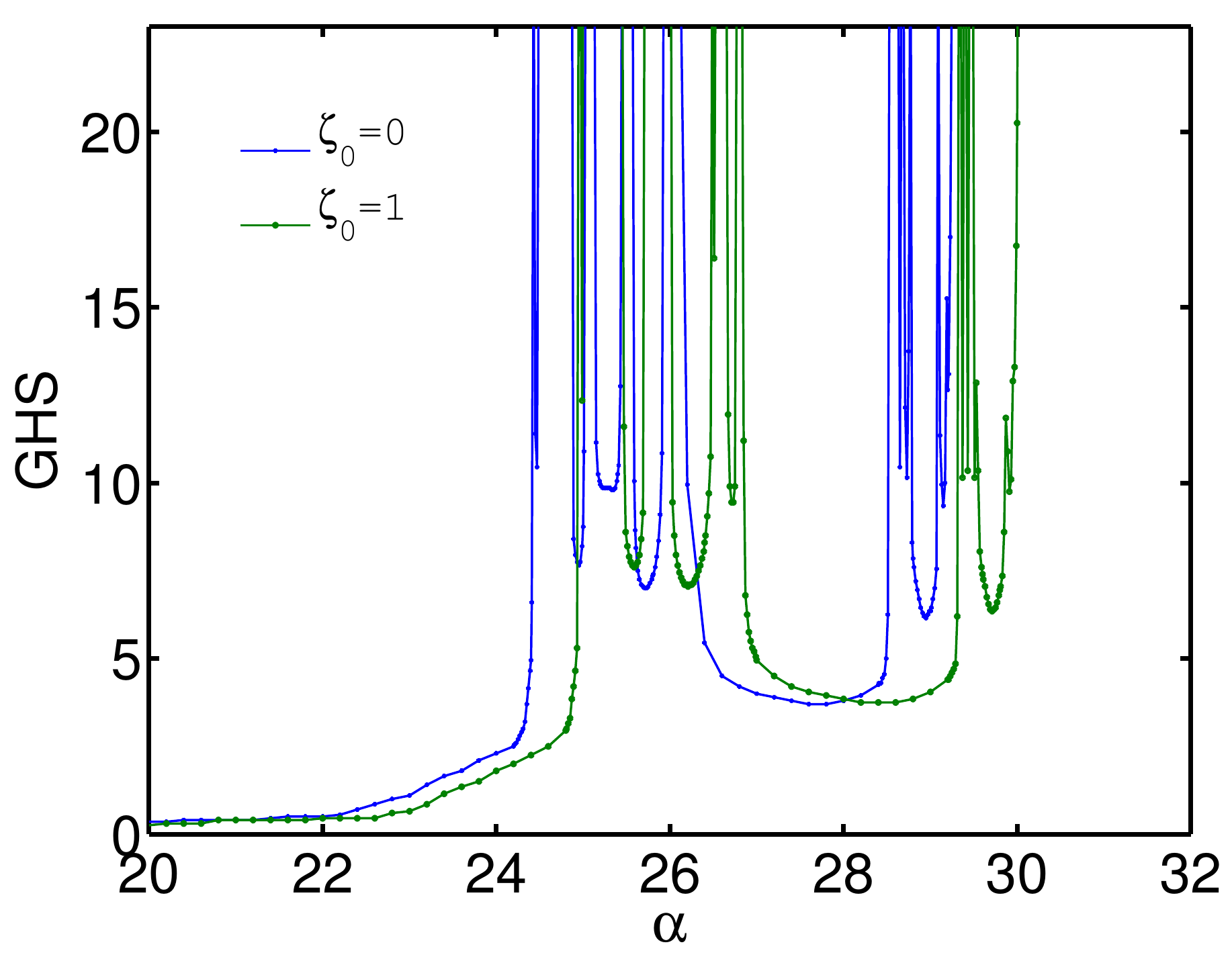}\\
(c)&(d)\\
\includegraphics[width=.5\columnwidth]{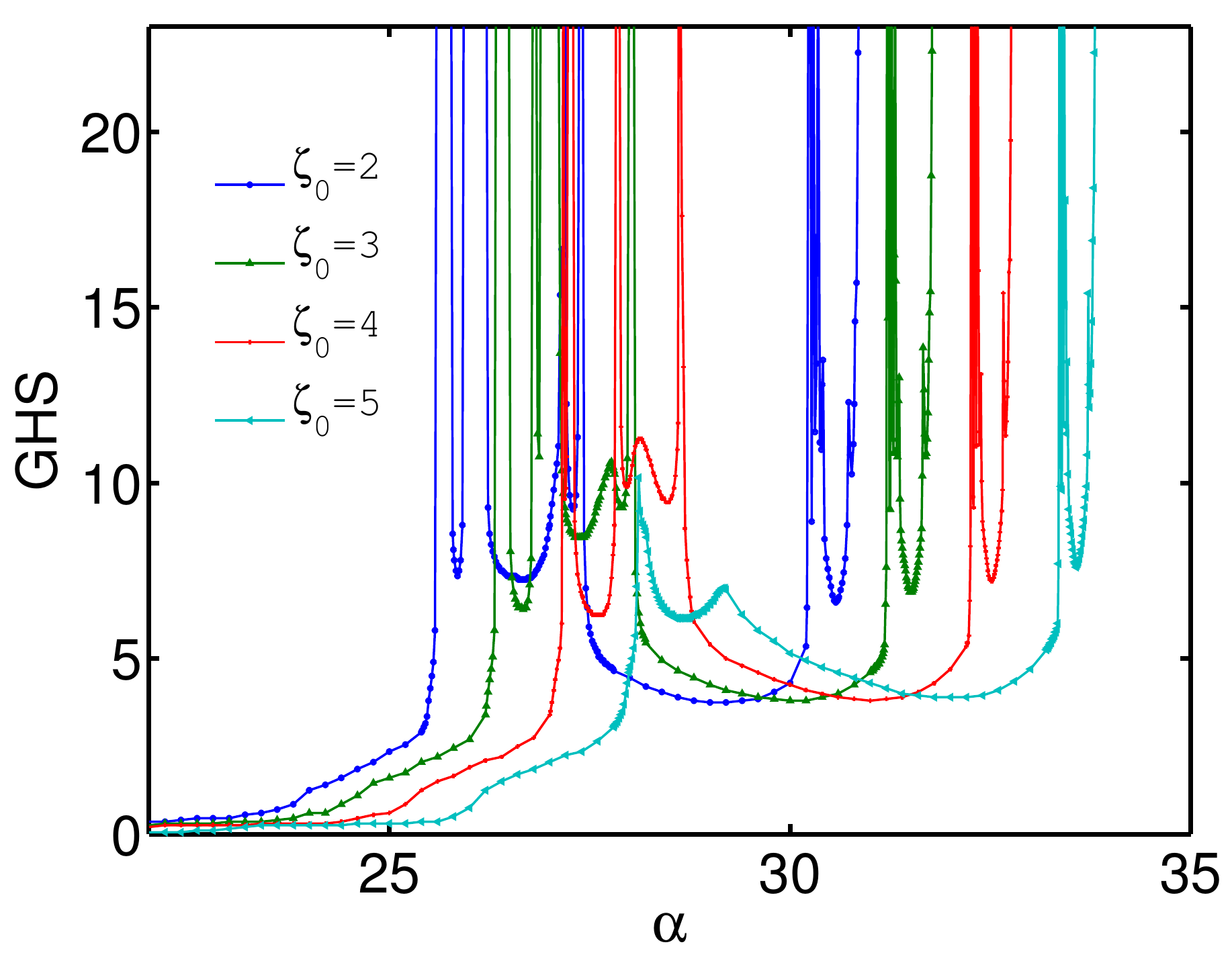}&
\includegraphics[width=.5\columnwidth]{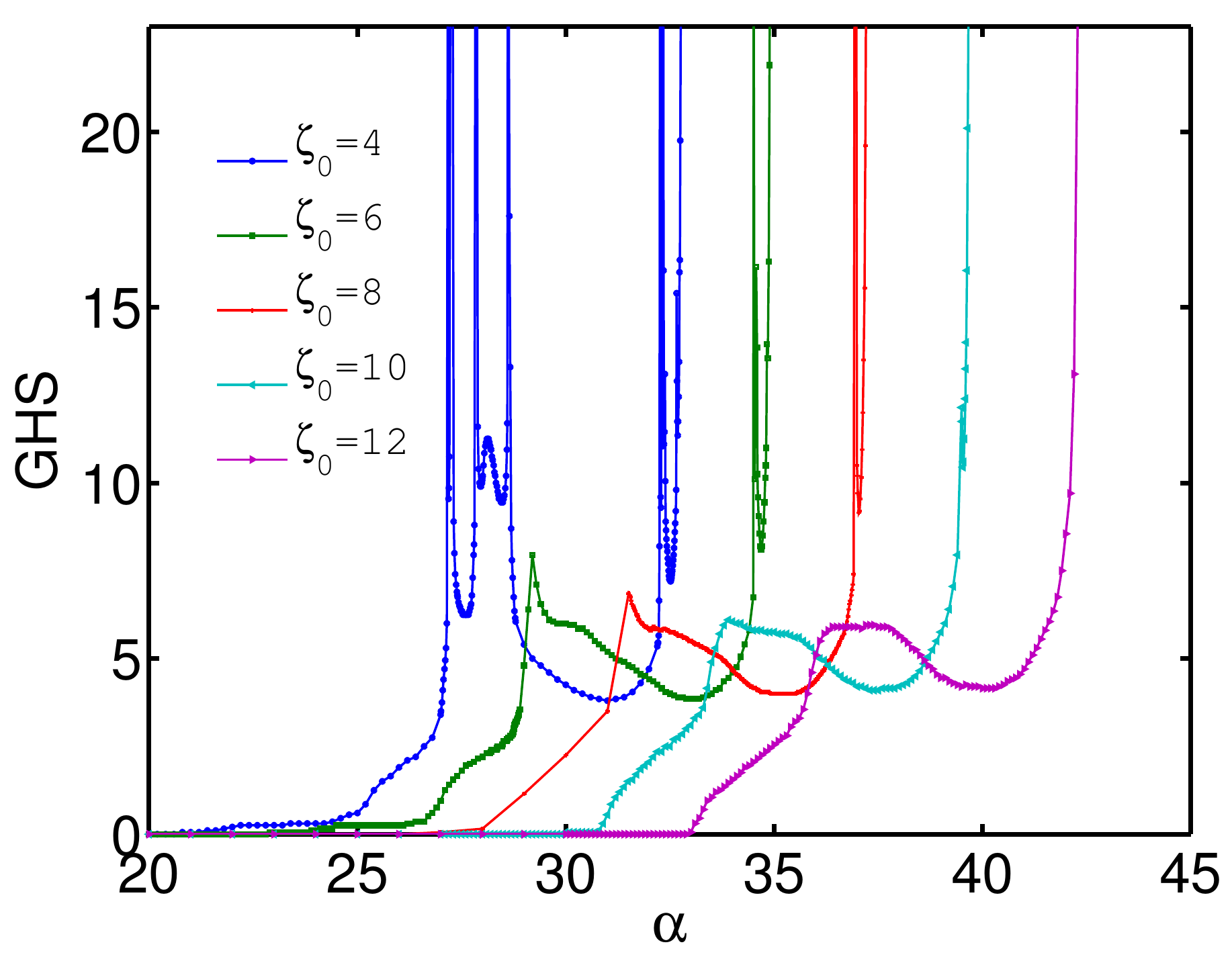}\\
(e)&(f)\\
\includegraphics[width=.5\columnwidth]{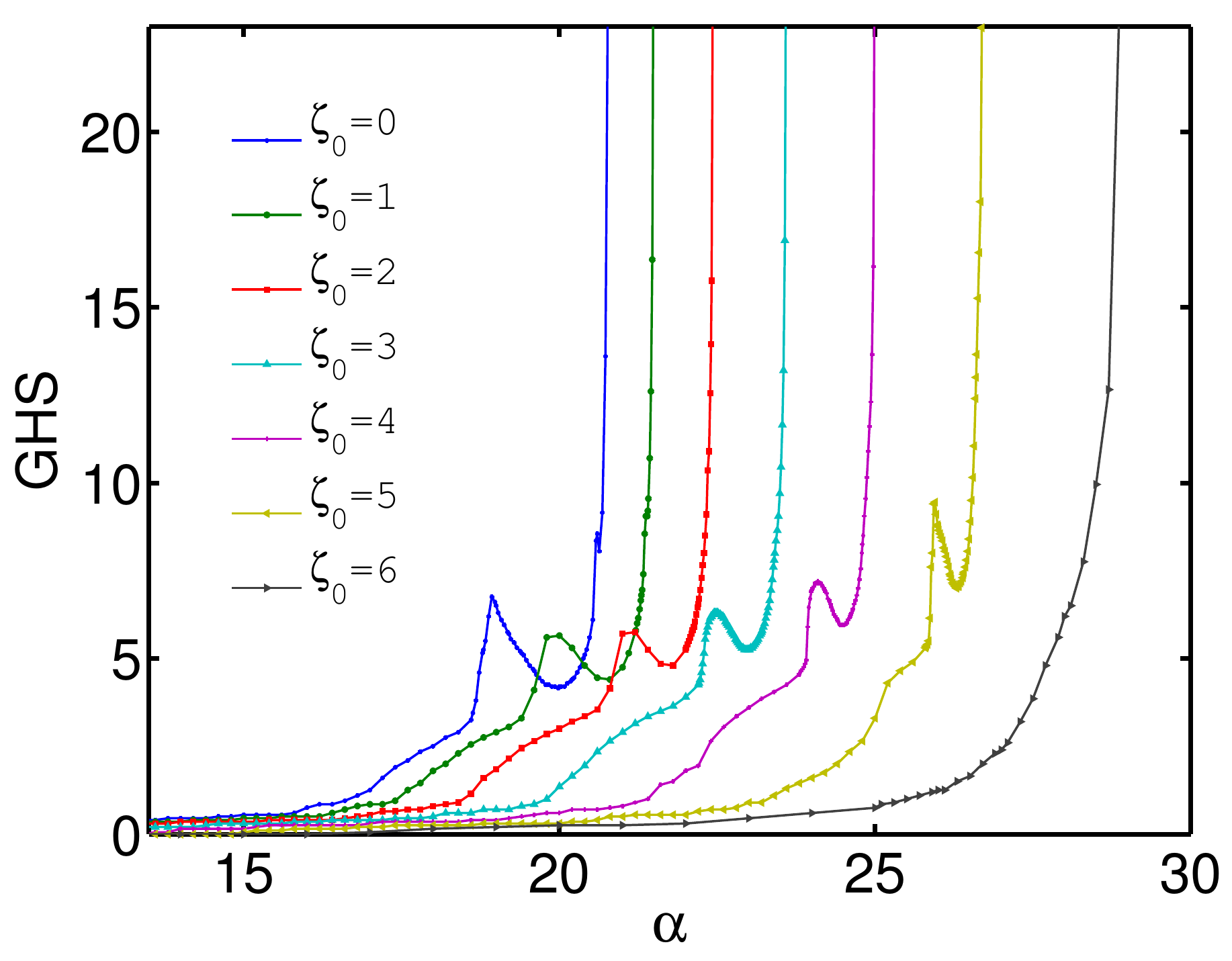}&
\includegraphics[width=.5\columnwidth]{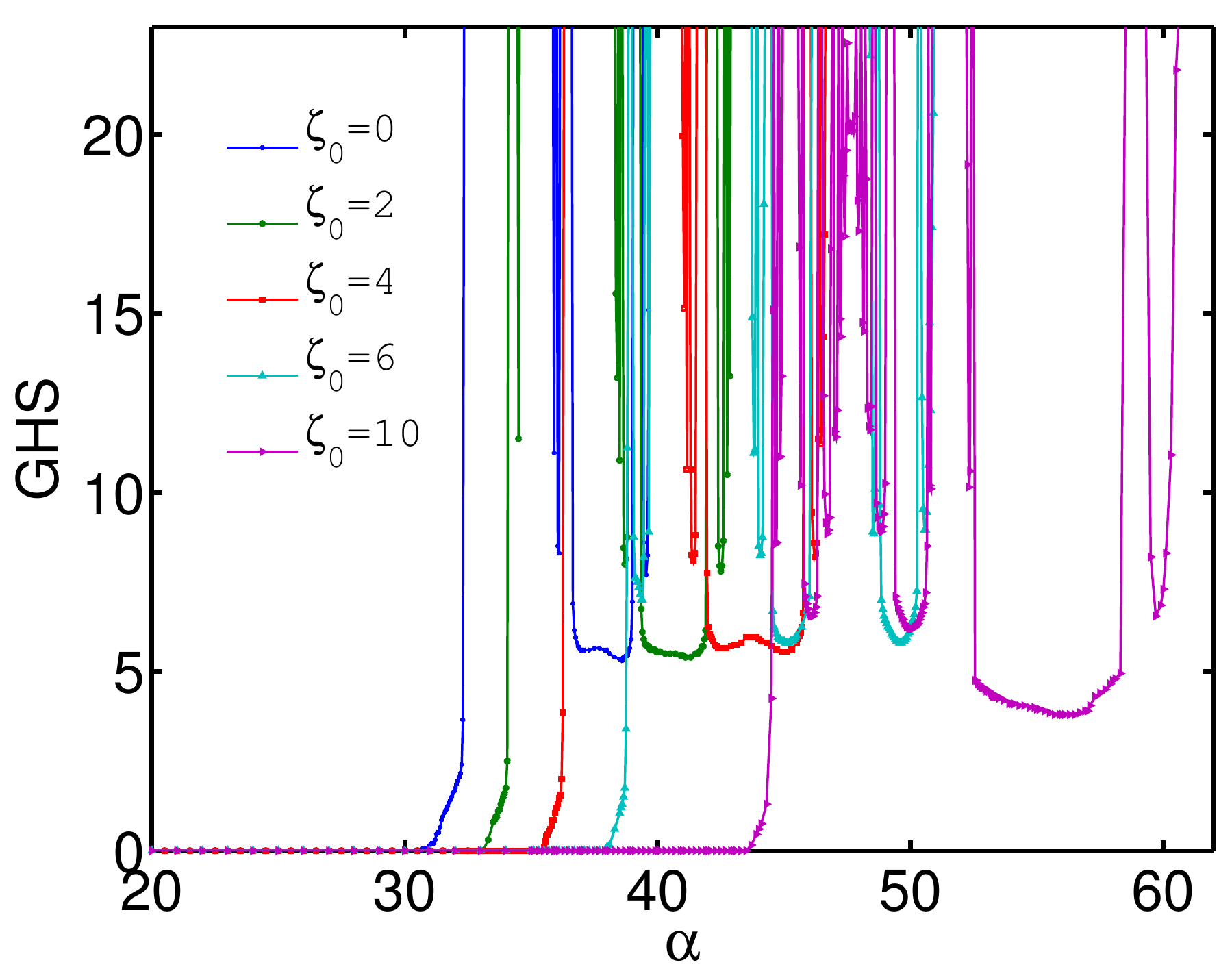}\\
\end{tabular}
\caption{GHS for different values of $\zeta_0$ and (a) $a=0.1$ $\theta=1^o$, (b) (c) and (d) $a=0.2$ $\theta=1^o$, (e) $a=0.2$ $\theta=2^o$ and (f) $a=0.3$ $\theta=1^o$.}\label{ghsa}
\end{figure}

\begin{figure}[htbp]
\centering
\begin{tabular}{cc}
(a)&(b)\\
\includegraphics[width=.5\columnwidth]{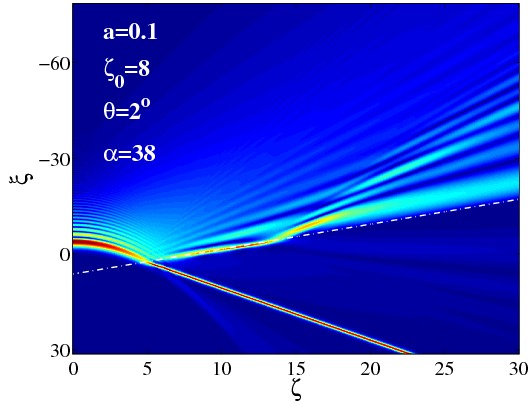}&
\includegraphics[width=.5\columnwidth]{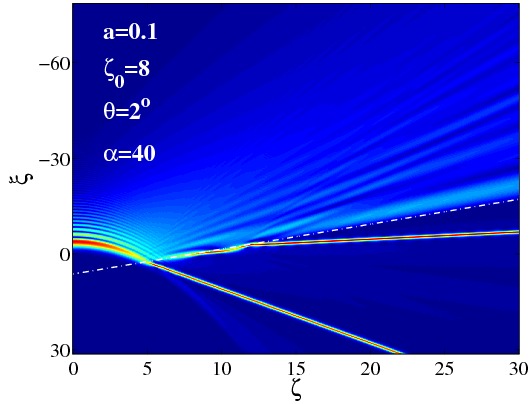}\\
(c)&(d)\\
\includegraphics[width=.5\columnwidth]{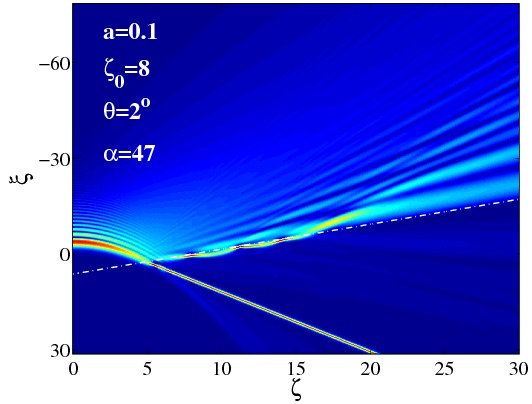}&
\includegraphics[width=.5\columnwidth]{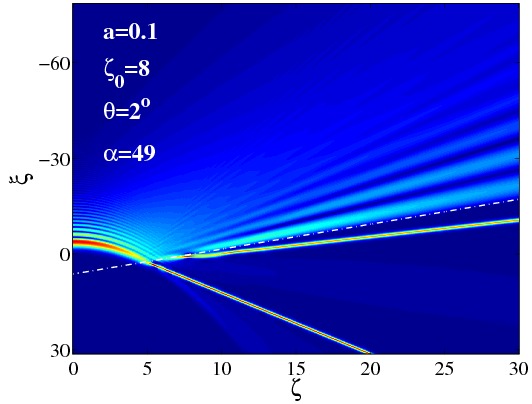}
\end{tabular}
\caption{Intermittent GHS sequence for the second emitted soliton.}\label{doble}
\end{figure}

We used the GHS, defined as the distance traveled by the peak of the beam in the second medium, versus the nonlinear mismatch $\alpha$ to analyze the results as shown in Figure \ref{ghsa}.  In this figure, the maximum GHS shown is limited by the total length of the computation window.   

Figure \ref{ghsa} (a) shows the GHS for $a=0.1$ at $\theta=1^o$ and $\zeta_0$ ranging from $0$ to $2$.  At the larger values of $\zeta_0$, and for any value of $\zeta_0$ with $\theta=2^o$, the process of generation of a soliton from the refracted beam dominates and GHS is not further observed.  Oscillatory GHS bands can be clearly identified in this figure for $\zeta_0=0$ and $2$.

Figures \ref{ghsa} (b), (c) and (d) show the results corresponding to $a=0.2$ at $\theta=1^o$.  Intermittency is clearly observed in all the cases with $\zeta_0\le8$, whereas oscillatory results are found for larger values. For $\zeta_0\le 4$ one finds two distinct bands of GHS intermittency separated by a region with small GHS.  For larger values of $\zeta_0$ the GHS enhancement in the band at smaller values of $\alpha$ is still responsible for oscillatory GHS even though intermittency is not observed in this band.  
When $a=0.2$ and $\theta=2^o$ (Figure \ref{ghsa} (e)) oscillatory GHS is found for values of $\zeta_0$ up to 5 and the oscillations cease at $\zeta_0=6$.  The soliton shedding process dominates at larger values.  

Figure  \ref{ghsa} (f) shows the gian GHS for $a=0.3$ and $\theta=1^o$:  different intermittency bands are found for all values of $\zeta_0$.  When $\theta=2^o$, intermittent GHS is found for $\zeta_0=0$ and $1$ and giant GHS with oscillations with $\alpha$ for $\zeta_0$ ranging from $2$ to $5$.  If $\zeta_0$ is further increased, after an oscillatory region the critically coupling to a surface mode is superseded by soliton shedding phenomena. 

When $a=0.4$ giant GHS with intermittency is observed when $\theta=1^o$ for all values of $\zeta_0$.  For $a=0.4$ and $\theta=2^o$, intermittent giant GHS is found when $\zeta_0$ is smaller or equal to 5 and oscillatory GHS at $\zeta_0=6$. For larger values, the behavior is similar to that of $a=0.3$ and $\theta=2^o$ at the largest values of $\zeta_0$. 

In general, one can observe how the decrease of the peak intensity of the finite energy input Airy beams  due to diffraction tends to shift the critical coupling features to larger values of $\alpha$ as the interface is moved to larger $\zeta_0$.   Even though the results for larger values of $a$ typically resemble those corresponding to an input Gaussian condition, quite remarkably, intermittency is also found for $a=0.6$, $\theta=2^o$ and $\zeta_0\ge8$ at very large values of $\alpha$.
 
At larger values of $\alpha$, one can find the emission of a second soliton in the nonlinear medium.  This secondary emitted wave can experience also a critical coupling to a surface mode and oscillatory and/or intermittent giant GHS, as shown in Figure \ref{doble}.

We have also studied some parameter ranges using the NLS obtained when the first term in \eqref{eq:nnse} is neglected with the result that the same qualitative behaviour is observed with significant quantitative differences.

In conclusion, we have studied the critical coupling to a nonlinear interface of finite energy Airy beams.  For very low values of the apodization parameter the beam curvature makes it difficult to define a grazing incidence condition and for very strong apodizations the response is close to that of a Gaussian beam.  In the intermediate regime we observe giant Goos-H\"anchen shifts that are intermittent or oscillatory as the input intensity varies.  This is markedly different from the results obtained when the input condition corresponds to a Gaussian beam or an optical soliton and it is a result of the larger complexity of the structure of the incident beam.  The observed effects can be interpreted as originating from the competition between soliton shedding from the refracted part of the beam and the critical coupling to a surface mode.  A very high sensitivity to the transmitted intensity is found for certain parameter regions and we have suggested the application of the new phenomena for optical sensing.   A detailed numerical study has been conducted to quantify these effects using a nonparaxial beam propagation scheme.  

This work has been funded by the Spanish MICINN, project number TEC2012-21303-C04-04, and Junta de Castilla y Le\'on, project number VA300A12-1.


\begin{thebibliography}{99}
%
%
\bibitem{berry} M.V. Berry and N.L. Balazs, Nonspreading wave packets, Am. J. Phys. {\bf 47,} 264--267 (1979). 

\bibitem{healing} J. Broky, G. A. Siviloglou, A. Dogariu, and D.N. Christodoulides, Self-healing properties of optical Airy beams, Opt. Express {\bf 16,} 12880--12891 (2008). 


\bibitem{gu} Y. Gu and G. Gbur, Scintillation of Airy beam arrays in atmospheric turbulence, Opt. Lett. {\bf 35,} 3456--3458
\bibitem{polynkin} P. Polynkin, M. Kolesik, J.V. Moloney, G.A. Siviloglou and D.N. Christodoulides, Curved Plasma Channel Generation Using Ultraintense Airy Beams, Science, {\bf 324,} 229-232 (2009).
\bibitem{siviloglou} G.A. Siviloglou and D. N. Christodoulides, Accelerating finite energy Airy beams, Opt. Lett. {\bf 32,} 979--981 (2007).
\bibitem{chremmos} I. D. Chremmos and N.K. Efremidis, Reflection and refraction of an Airy beam at a dielectric interface, J. Opt. Soc. Am. B {\bf 29,} 861--868 (2012).
\bibitem{goos47} F. Goos and H. H\"anchen, Ann. Phys {\bf 1,} 333 (1947).
\bibitem{tomlinson82} W.J. Tomlinson, J.P. Gordon, P.W. Smith and A.E. Kaplan, Reflection of a Gaussian beam at a nonlinear interface, Appl. Opt. {\bf 21,} 2041--2051 (1982). 
\bibitem{aceves} A.B. Aceves, J.V. Moloney and A.C. Newell, Theory of light-beam propagation at nonlinear interfaces. I. Equivalent-particle theory for a single interface, Phys. Rev. A {\bf 39,} 1809 (1989).
\bibitem{chamorro98} P. Chamorro-Posada, G.S. McDonald and G.H.C. New, Nonparaxial solitons, J. Mod. Opt. {\bf 45,}1111-1121 (1998) .
\bibitem{sanchez07} J. S\'anchez-Curto, P.Chamorro-Posada and G.S. McDonald, Helmholtz soltions at nonlinear interfaces,  Opt. Lett. {\bf 32,}  1126--1128  (2007).
\bibitem{chamorro01} P. Chamorro-Posada, G.S. McDonald and G.H.C. New, Nonparaxial beam propagation methods, Opt. Commun. {\bf 192,} 1-12 (2001). 
\bibitem{tlm} P. Chamorro-Posada and G.S. McDonald, Time-domain analysis of Helmholtz soliton propagation usign the TLM method, J. Nonlinear Opt. Phys. Mat. {\bf 21,} 1250031 (2012).

\bibitem{sanchez11} J. S\'anchez-Curto, P. Chamorro-Posada and G.S. McDonald , Giant Goos-Hanchen shifts and radiation-induced trapping of Helmholtz solitons at nonlinear interfaces, Opt. Lett. {\bf 36,}  3605-3607 (2011).

\bibitem{arxiv} P. Chamorro-Posada, J. S\'anchez-Curto, A.B. Aceves and G.S. McDonald, On the asymptotic evolution of finite energy Airy wavefunctions, arXiv:1305.3529
\bibitem{fattal2011} Y. Fattal, A. Rudnick A and D.M Marom, Soliton shedding from Airy pulses in Kerr media, Opt. Express {\bf 19,} 17298-307 (2011).
\bibitem{tamm} H. Maaba, M. Faryadb and A. Lakhtakiab, Prism-coupled excitation of multiple Tamm waves, J. Mod. Opt. {\bf 60} 355–358 (2013).
\end{thebibliography}
\end{document}